\documentclass[twocolumn]{aastex631}
\usepackage{natbib}
\usepackage{graphicx}
\usepackage{mathrsfs}
\usepackage{amsmath}
\usepackage{hyperref}
\usepackage{enumitem}

\begin{document}

\title{The Carbon-Deficient Evolution of TRAPPIST-1c}

\author{Katie E. Teixeira}
\author{Caroline V. Morley}
\affiliation{Department of Astronomy, University of Texas at Austin, Austin, TX, USA}
\affiliation{Center for Planetary Systems Habitability, University of Texas, Austin, TX, USA}
\author{Bradford J. Foley}
\affiliation{Department of Geosciences, Pennsylvania State University, University Park, PA, USA}
\author{Cayman T. Unterborn}
\affiliation{Space Instruments and Payload Department, Southwest Research Institute, San Antonio, TX, USA}

\begin{abstract}
Transiting planets orbiting M dwarfs provide the best opportunity to study the atmospheres of rocky planets with current facilities. As \textit{JWST} enters its second year of science operations, an important initial endeavor is to determine whether these rocky planets have atmospheres at all. M dwarf host stars are thought to pose a major threat to planetary atmospheres due to their high magnetic activity over several billion-year timescales, and might completely strip atmospheres. Several Cycle 1 and 2 GO and GTO programs are testing this hypothesis, observing a series of rocky planets to determine whether they retained their atmospheres. A key case-study is TRAPPIST-1c, which receives almost the same bolometric flux as Venus. We might, therefore, expect TRAPPIST-1c to possess a thick, $\mathrm{CO}_2$-dominated atmosphere. Instead, \citet{Zieba_23} show that TRAPPIST-1c has little to no CO$_2$ in its atmosphere. To interpret these results, we run coupled time-dependent simulations of planetary outgassing and atmospheric escape to model the evolution of TRAPPIST-1c’s atmosphere. We find that the stellar wind stripping that is expected to occur on TRAPPIST-1c over its lifetime can only remove up to $\sim 16$ bar of CO$_2$, less than the modern $\mathrm{CO}_2$ inventory of either Earth or Venus. Therefore, we infer that TRAPPIST-1c either formed volatile-poor, as compared to Earth and Venus, or lost a substantial amount of $\mathrm{CO}_2$ during an early phase of hydrodynamic hydrogen escape. Finally, we scale our results for the other TRAPPIST-1 planets, finding that the more distant TRAPPIST-1 planets may readily retain atmospheres. 
\end{abstract}


\section{Introduction}

Over 5000 exoplanets have now been discovered, the majority of them found through transits with NASA's \textit{Kepler} and \textit{TESS} missions. About 30\% of known exoplanets have radii less than 2 $R_\oplus$, classifying them as possibly rocky \citep{Kaltenegger_17,Fulton_17}. While it has been possible to constrain the radii and sometimes the masses of these potentially rocky worlds, it has not been possible, until recently, to characterize their atmospheres. While the atmospheres of rocky planets around the smallest stars may be detectable with \textit{JWST} \citep{Morley_17, Lustig-Yaeger_19}, there remains great uncertainty in whether planets orbiting these small stars will retain their atmospheres for billions of years \citep{Lammer_07, Tian_09, Cohen_15, Bourrier_17, Garraffo_17, Becker_20, France_20, Krissansen-Totton_22, doAmaral_22}. 

Prior observations of rocky planets orbiting late M dwarfs have allowed us to empirically rule out the presence of thick atmospheres around three such planets. \textit{Spitzer}, along with data from \textit{TESS} and \textit{K2}, allowed us to obtain phases curves for LHS 3844b \citep{Kreidberg_19} and K2-141b \citep{Zieba_22} and secondary eclipse photometry for GJ 1251b \citep{Crossfield_22}. Each of these planets were found to have dayside temperatures consistent with that of a bare rock with no thick atmosphere redistributing heat to the nightside. As of mid-2023, \textit{JWST} has completed its first year of science operations, and we are now observing a series of terrestrial planets orbiting nearby M dwarfs. Among the first observations published is a transmission spectrum of LHS 475b, which is flat, consistent with no atmosphere or a high-altitude cloud \citep{Lustig-Yaeger_23}. Another such observation of the transmission spectrum of GJ 486b \citep{Moran_23} has found a feature in the spectrum, which is consistent with either a water-rich atmosphere or stellar contamination on a planet orbiting an M3.5 dwarf. 

Of particular interest to the exoplanet community is the M8V star TRAPPIST-1 \citep{Gillon_17}. Its seven transiting planets occupy locations which are interior to, within, and outside of the habitable zone. This makes it a case study in which to probe the distance from the host star where atmospheres are retained. This is important because atmospheres are taken to be a prerequisite for life. All of the TRAPPIST-1 planets will be observed by JWST in its first two Cycles of operations, with a mix of transit observations (all planets) and eclipse observations (planets b and c). 
The first \textit{JWST} eclipse observations of the TRAPPIST-1 system have already shown that the innermost planet, TRAPPIST-1b, has no thick $\mathrm{CO}_2$ atmosphere \citep{Greene_23}, with models from \citet{Ih_23} ruling out pure CO$_2$ atmospheres thicker than 6.5 mbar and atmospheres containing 100 ppm CO$_2$ thicker than 0.3 bar. 

\subsection{Atmospheric Escape from Terrestrial Planets}
It is currently unknown whether small M dwarfs like TRAPPIST-1 are hospitable to atmospheres at any orbital distance. M dwarfs can be highly magnetically active for up to 6-7 billion years, significantly longer than Sun-like stars that diminish in activity by 1 billion years \citep{West_06}. Magnetic activity presents itself as high-energy X-ray and ultraviolet (XUV) radiation, flares, winds, and coronal mass ejections, all of which could be harmful to planetary atmospheres.

In the Solar System, there is evidence that the terrestrial planets formed with primordial atmospheres of hydrogen and helium which were mostly lost as the gas of the protoplanetary disk dissipated \citep{Lammer_18, Young_23}. Secondary atmospheres made up largely of volatiles such as carbon dioxide ($\mathrm{CO}_2$) and water ($\mathrm{H}_2\mathrm{O}$) then formed by planetary outgassing as magma oceans solidified. Earth was able maintain at least part of its initial water reservoir during this hot magma ocean phase, and the geological carbon cycle (i.e., carbon-silicate cycle) has subsequently recycled volatiles into and out of the atmosphere, regulating the planet's climate. 

Conversely, Venus may be representative of planets which become desiccated during the magma ocean phase, because they receive too much stellar flux for their steam atmospheres to cool efficiently, triggering a runaway greenhouse \citep{Hamano_13}. Water is then lost to space, by its rapid photodissociation and subsequent hydrodynamic escape of hydrogen \citep{Kasting_88}. This desiccation left Venus with the $\mathrm{CO}_2$-dominated atmosphere that we see today. Mars also has a $\mathrm{CO}_2$-dominated atmosphere, albeit much thinner, but had a warmer, wetter atmosphere in the past, most of which has been lost. These atmospheres are not static, even today, as modern probes detect the steady loss of atmospheric material from Earth, Venus and Mars, mostly in the form of ion escape by solar wind \citep{Jakosky_18, Edberg_11}. In fact, escape has affected their atmospheres over Gyr timescales \citep{Lammer_08}. 

The history of the Solar System terrestrial planets can inform our understanding of how the atmospheres of terrestrial exoplanets evolve, specifically close-in planets like TRAPPIST-1c, which is the subject of this paper. For planets that orbit close enough to their host star to evaporate water into the atmosphere, a period of $\mathrm{H}_2\mathrm{O}$ photo-dissociation and intense early hydrodynamic loss of hydrogen (which results in net water loss) is expected. Hydrodynamic escape occurs when the light constituent of an atmosphere reaches a high enough temperature to leave in an evaporative wind. During this hydrodynamic phase, it is also possible for heavy constituents to be dragged along via the hydrodynamic wind. Therefore, a net loss of heavier constituents like $\mathrm{CO}_2$ and $\mathrm{O}_2$ can occur. Once the hydrodynamic phase is over, we expect the remaining atmosphere to be dominated by heavier molecules such as $\mathrm{CO}_2$, as in the case of Venus.

Now, over the rest of the planet's lifetime, $\mathrm{CO}_2$ can be lost slowly through other mechanisms. Jeans escape, another type of thermal escape, is a mechanism that occurs when the high-velocity, high-altitude particles of a gas in thermodynamic equilibrium have velocities which exceed the escape velocity of the planet. Jeans escape is not dominant in a $\mathrm{CO}_2$-dominated atmosphere, because $\mathrm{CO}_2$ is too heavy and cools efficiently through the 15 $\mu m$ band \citep{Gordiets_85}. Therefore, we expect that long-term escape on these types of planets is dominated by non-thermal escape, particularly in the form of stellar wind stripping as seen in the Solar System.

Even as stellar phenemona strip material from planetary atmospheres, terrestrial planet atmospheres are replenished by geological processes, such as volcanic outgassing, as seen in recent geological history on Earth, Venus and Mars. This outgassing may not last for the entire lifetime of the planet, as \cite{Unterborn_22} suggests for the TRAPPIST-1 planets. Nevertheless, we can model both the escape and outgassing of terrestrial planet atmospheres as coupled time-dependent processes that evolve an atmosphere through time. 

\subsection{This Work in Context}

In this paper, we use coupled, time-dependent simulations of outgassing and escape to model the terrestrial planet TRAPPIST-1c. Because TRAPPIST-1c orbits close to its host star, receiving slightly more flux than Venus, we assume here that it has lost effectively all $\mathrm{H}_2\mathrm{O}$ and possesses a Venus-like $\mathrm{CO}_2$-dominated atmosphere, as is common in the Solar System. Recent observations have been taken for TRAPPIST-1c, using \textit{JWST}'s MIRI instrument to obtain secondary eclipse photometry in the 15 $\mu m$ $\mathrm{CO}_2$ absorption band. The data from these observations \citep{Zieba_23} point to the absence of a thick $\mathrm{CO}_2$ atmosphere, ruling out pure $\mathrm{CO}_2$ atmospheres with surface pressures $\geq 0.1$ bar. \cite{Zieba_23} find that a bare rock surface or a thin $\mathrm{O}_2$-dominated atmosphere could also be consistent with their data. \cite{Lincowski_23} further explored the potential compositions of TRAPPIST-1c's present atmosphere. They acknowledge that $\mathrm{CO}_2$ is likely to be a principal component of outgassing for Earth-sized or larger planets \citep{Gaillard_14}, but test many different atmospheric species in their models. They confirm the results of \cite{Zieba_23} but additionally find that a maximum of 10$\%$ $\mathrm{H}_2\mathrm{O}$ abundance is consistent with the observations.

Assuming $\mathrm{CO}_2$ has dominated TRAPPIST-1c's atmosphere for most of its history, the constraint on $\mathrm{CO}_2$ surface pressure from \cite{Zieba_23} allows us to constrain the history of its atmosphere with simulations and assess its early volatile inventory. We base our simulation framework and study upon the work of \cite{Kane_20} which found volatile-poor formation conditions for LHS 3844b. Our simulations differ from \cite{Krissansen-Totton_22} which made predictions for the atmospheres of TRAPPIST-1 planets. \cite{Krissansen-Totton_22} includes a range of atmospheric species, including $\mathrm{H}_2\mathrm{O}$, $\mathrm{CO}_2$, and $\mathrm{O}_2$, and runs coupled simulations starting in the magma ocean phase all the way to present day. We first calculate the amount of $\mathrm{CO}_2$ potentially lost during the hydrodynamic escape phase, and then run coupled time-dependent simulations of outgassing and escape of $\mathrm{CO}_2$ starting after magma ocean solidification and atmospheric desiccation. Our simulations, therefore, track the amount of $\mathrm{CO}_2$ lost in the long-term stellar wind stripping phase of TRAPPIST-1c's evolution. This allows us to determine whether long-term stellar wind stripping is efficient enough to chisel the atmosphere down to what we see today. If not, formation conditions or significant early escape must be responsible for the current shortage of $\mathrm{CO}_2$ in TRAPPIST-1c's atmosphere.

We organize this manuscript as follows. In Section \ref{sec:Methods}, we describe the components of our simulations of TRAPPIST-1c, including stellar properties, planet properties, planetary outgassing, and atmospheric escape. We detail how we calculate outgassing rates from interior thermal evolution and escape rates from stellar-wind-induced ion escape. We explain how we couple these processes in time-dependent simulations and vary key parameters to create a grid of potential atmospheric histories of TRAPPIST-1c. In Section \ref{sec:Results}, we first provide calculations of $\mathrm{CO}_2$ loss during the early hydrodynamic escape phase, and then we analyze the results of our long-term simulations, identifying the parameters of those that fit the observational constraint. In Section \ref{sec:Discussion}, we discuss the implications of these results for the TRAPPIST-1 system as a whole. We address caveats of our models, diagnosing future developments that will help us better characterize the atmosphere of TRAPPIST-1c and its neighbors. In Section \ref{sec:Conclusions}, we summarize our findings and conclude.

\begin{figure*}
    \centering
    \includegraphics[width=6.8in]{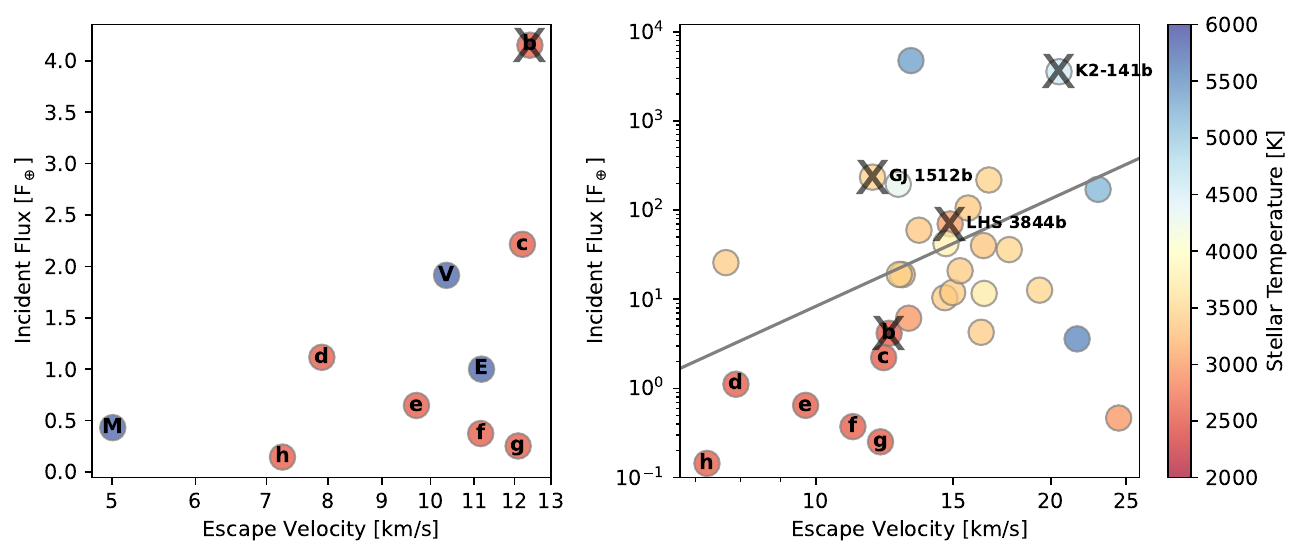}
    \caption{Comparing TRAPPIST-1 planets to Solar System planets (left) and likely rocky exoplanets already observed or planned to be observed by \textit{JWST} in Cycle 1 and 2 (right). Incident flux is plotted against escape velocity, so that planets in the top left are less likely to retain atmospheres, and planets in the bottom right are more likely to retain atmospheres. Stellar temperature, represented by color, may also play a role in atmosphere retention. Black Xs are placed on the labeled exoplanets which have already been observed to have no thick atmosphere. The gray line in the right panel is the ``cosmic shoreline'' ($\mathrm{F_{inc}} \propto \mathrm{v_{esc}}^4$) presented in \cite{Zahnle_17}.}
    \label{fig:Planet-Context}
\end{figure*}

\section{Methods}
\label{sec:Methods}

We model the atmosphere of TRAPPIST-1c as a coupled system that includes the interior of the planet and the stellar environment. We encode observationally-constrained properties of TRAPPIST-1c and its host star into simulations which evolve the atmosphere through the planet's lifetime. In our simulations, planetary outgassing delivers $\mathrm{CO}_2$ to the atmosphere and is based on geophysical phenomena observed on Earth, except without plate tectonics, as on present-day Venus. Atmospheric escape removes $\mathrm{CO}_2$ from the atmosphere and is based on the stellar-wind-induced ion escape of Venus and Mars observed and modeled by magnetohydrodynamic (MHD) simulations, accounting for the differing properties of M dwarfs like TRAPPIST-1. Coupling these processes then allows us to track the amount of $\mathrm{CO}_2$ in the atmosphere at any given time.

\subsection{Stellar Properties}
\label{sec:Stellar-Properties}
We initialize our simulations with the properties of the M8V star TRAPPIST-1 \citep{Costa06}. According to \cite{Agol21}, TRAPPIST-1 has mass $M_\star=0.0898\pm0.0023 \ M_\odot$, radius $R_\star=0.1192\pm0.0013 \ R_\odot$, effective temperature $T_\mathrm{eff}=2,566\pm26$ K, and bolometric luminosity $L_\mathrm{bol}=0.000553\pm0.000018 \ L_\odot$.

We use the age of TRAPPIST-1 presented in \cite{Burgasser_17}. They conclude that TRAPPIST-1 has an age of $7.6\pm2.2$ Gyr. The rotation period of TRAPPIST-1 is $3.30\pm0.14$ days as found photometrically by \cite{Luger_17}. Age and rotation period are theoretically related to each other and to magnetic activity which we address further in Section \ref{sec:Atmospheric-Evolution} and Section \ref{sec:Star-Planet-Interaction}.

\newpage
\subsection{Planet Properties}
\label{sec:Planet-Properties}

In this manuscript, we focus on the second-closest planet to the host star, TRAPPIST-1c. TRAPPIST-1c has mass $M_p=1.308\pm0.056 \ M_\oplus$, radius $R_p=1.097\substack{+0.014 \\ -0.012}\ R_\oplus$, and semi-major axis $a=0.01580\pm0.00013$ AU \citep{Agol21}. These planetary properties are displayed with the stellar properties in Table \ref{table:Properties}. Mass and radius constraints provide an estimate for the density of TRAPPIST-1c, approximately $\rho=5.45 \ \mathrm{g}/\mathrm{cm}^3$, which is almost exactly Earth-like. The surface gravity of TRAPPIST-1c is $10.65\pm0.42 \ \mathrm{m}/\mathrm{s}^2$, slightly larger than Earth's. Given the semi-major axis and stellar luminosity, we calculate the incident bolometric flux on TRAPPIST-1c to be approximately $F_\mathrm{inc}=2.22F_{\mathrm{inc},\oplus}$, which is near the value for Venus ($1.91F_{\mathrm{inc},\oplus}$).

\begin{table}[ht!]
\centering
 \begin{tabular}{l l}
 $\mathrm{Parameter}$ & $\mathrm{Value}$ \\
 \hline\hline
 \textbf{Star - TRAPPIST-1} \\
 Mass $M_\star$ & $0.0898\pm0.0023 \ M_\odot$ \\
 Radius $R_\star$ & $0.1192\pm0.0013 \ R_\odot$ \\
 Effective temperature $T_\mathrm{eff}$ & $2566\pm26$ K \\
 Bolometric luminosity $L_\mathrm{bol}$ & $0.000553\pm0.000018 \ L_\odot$ \\
 Age & $7.6\pm2.2$ Gyr \\
 Rotation period $P_{\mathrm{rot}}$ & $3.30\pm0.14$ days \\
 \textbf{Planet - TRAPPIST-1c} \\
 Mass $M_p$ & $1.308\pm0.056 \ M_\oplus$ \\
 Radius $R_p$ & $1.097\substack{+0.014 \\ -0.012}\ R_\oplus$ \\
 Semi-major axis $a$ & $0.01580\pm0.00013$ AU \\
 \hline
\end{tabular}
\caption{Relevant properties of TRAPPIST-1 and TRAPPIST-1c as cited in the text.}
\label{table:Properties}
\end{table}

The aforementioned properties of TRAPPIST-1c are compared to that of its neighbors, Solar System planets, and other likely rocky exoplanets in Figure \ref{fig:Planet-Context}. The left panel displays the incident flux, escape velocity, and stellar temperature for the TRAPPIST-1 planets, Earth, Venus, and Mars. The right panel displays the same information for the TRAPPIST-1 planets and other exoplanets whose atmospheres have been observed, or will be observed in JWST Cycle 1 and Cycle 2 GO and GTO programs. Black Xs are placed on the exoplanets, LHS 3844b, GJ 1252b, K2-141b, and TRAPPIST-1b, which have been observed to have no thick atmosphere \citep{Kreidberg_19,Crossfield_22,Zieba_22,Greene_23}. It is plausible that the existence of atmospheres has some dependence on all three variables. The closer a planet is to its host star, the more high-energy radiation and stellar material bombardment it is expected to receive. Higher escape velocity makes it less likely that atmospheric material will escape at a given temperature and mean molecular mass. Finally, atmospheres may be favored in higher stellar temperature cases where high-activity pre-main sequence phases are shorter. The gray line in the right panel is the ``cosmic shoreline'' ($\mathrm{F_{inc}} \propto \mathrm{v_{esc}}^4$) presented in \cite{Zahnle_17}, and is an empirical division of planets known with and without atmospheres. This dividing line might suggest that the TRAPPIST-1 planets should all have atmospheres, but that appears not to be the case. The exact effect that incident flux, escape velocity, and stellar temperature have on the presence of atmospheres is, thus, still a lingering question and must be explored through observations and simulations.

\subsection{Atmospheric Evolution}
\label{sec:Atmospheric-Evolution}

In our simulations, the $\mathrm{CO}_2$ atmosphere of TRAPPIST-1c evolves through two mechanisms: planetary outgassing and atmospheric escape. Outgassing is a result of planet interior processes that are driven by mantle convection, thereby cooling the mantle over time. Tracking the mantle temperature over time along with several other important quantities allows outgassing rates to be calculated. Atmospheric escape is a result of many complex physical and chemical processes that we do not model directly here but we instead employ the results of magnetohydrodynamic (MHD) models that trace ion escape driven by the stellar wind. We discuss this choice further in Section \ref{sec:Discussion}.

\subsubsection{Planetary Outgassing}
\label{sec:Planetary-Outgassing}

We employ a new version of the thermal evolution model presented in \cite{Foley_18}, translated to Python and available at \url{https://github.com/katieteixeira/atmospheric_evolution} \citep{teixeira_&_foley_code}, to calculate outgassing rates. This thermal evolution code models Earth-like planets in the stagnant lid regime, where lithospheres are rigid and immobile, but volcanism still occurs. As far as we know, Earth is the only Solar System planet that has plate tectonics. Both Venus and Mars have stagnant lids, so we expect this tectonic mode to be common among terrestrial planets \citep{Foley_16}. The likely lack of liquid water on TRAPPIST-1c also disfavors plate tectonics, as water has been shown to be important for plate tectonics \citep{Hirth_96}. Water lowers the strength of the lithosphere, weakening plate boundaries and allowing subduction to occur \citep{Regenauer-Lieb_01, Korenaga_07}. 

The original code presented in \cite{Foley_18} couples the thermal evolution of stagnant lid planet interiors to melting, crustal growth, weathering, and $\mathrm{CO}_2$ outgassing. Convection transports heat and carbonated material to the base of the stagnant lid, where melting and eruption allow them to escape to the surface. On stagnant lid planets, this volcanism itself is responsible for crustal growth and the creation of weatherable rock. If carbon is weathered back into the crust, it can be buried and recycled either into the mantle, by carbon foundering, or the atmosphere, by metamorphic outgassing.

As in \cite{Foley_18}, the temperature of the mantle evolves according to the following equation:
\begin{equation}
\label{eq:Mantle temperature}
    V_{man} \rho c_p \frac{dT_p}{dt} = Q_{man} - A_{man}F_{man} - f_m \rho_m (c_p \Delta T_m + L_m)
\end{equation}
where $V_{man}$ is the volume of convecting mantle beneath the stagnant lid, $\rho$ is the bulk density of the mantle, $c_p$ is the heat capacity, $T_p$ is the potential temperature of the upper mantle, $t$ is time, $Q_{man}$ is the radiogenic heating rate in the mantle, $A_{man}$ is the surface area of the top of the convecting mantle, $F_{man}$ the heat flux from the convecting mantle at the base of the stagnant lid, $f_m$ is the volumetric melt production, $\rho_m$ is the density of the mantle melt, $L_m$ is the latent heat of fusion of the mantle, and $\Delta T_m$ is the temperature difference between the melt erupted at the surface and the surface temperature. This assumes that heat from the core does not play a role in mantle thermal evolution.

Coupled to Equation \ref{eq:Mantle temperature} and several other differential equations, the following equation calculates the rate at which carbon is outgassed from the mantle to the atmopshere:
\begin{equation}
\label{eq:Outgassing rate}
    \frac{dR_{man}}{dt} = -\frac{f_m R_{man} [1 - (1- \phi)^{1/D_{CO_2}}]}{\phi (V_{man} + V_{lid})}
\end{equation}
where $R_{man}$ is the mantle carbon reservoir, $\phi$ is the melt fraction, $D_{CO_2}$ is the distribution coefficient of $\mathrm{CO}_2$, and $V_{lid}$ is the volume of the stagnant lid. This assumes that a certain fraction of $\mathrm{CO}_2$ enters the melt phase and a fraction of the melt degasses its $\mathrm{CO}_2$ to the atmosphere. We refer the reader to \cite{Foley_18} for more details.

\begin{figure}
    \centering
    \includegraphics[width=3.4in]{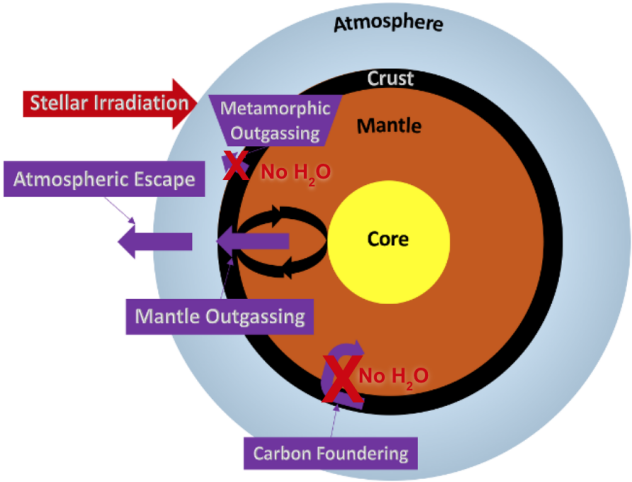}
    \caption{Processes that move atmospheric species into and out of a stagnant lid planet’s atmosphere. Without liquid water, metamorphic outgassing and carbon foundering do not occur.}
    \label{fig:Schematic}
\end{figure}

By default, the thermal evolution code assumes that water is present, and thus, weathering occurs. Since we assume TRAPPIST-1c has no liquid water on its surface, we ``turn off'' weathering in the code. No $\mathrm{CO}_2$ is returned to the crust, and, therefore, no $\mathrm{CO}_2$ is recycled into the mantle, so metamorphic outgassing and carbon foundering do not operate. This implies that all $\mathrm{CO}_2$ which is outgassed will simply build up in the atmosphere. This is captured in Figure \ref{fig:Schematic} with the major processes that potentially move $\mathrm{CO}_2$ into and out of TRAPPIST-1c's atmosphere.

\subsubsection{Atmospheric Escape}
\label{sec:Atmospheric-Escape}

We parameterize the atmospheric escape component of our simulations based on \cite{Dong_18}, which used the BATS-R-US MHD code \citep{vanderHolst_14} to model the stellar wind of TRAPPIST-1 and its interaction with the seven TRAPPIST-1 planets. BATS-R-US has been tested and validated on Solar System planets and is based on the composition of Mars and Venus (e.g. \citealt{Ma_04,Ma_13}). To obtain an upper bound on escape rates, the TRAPPIST-1 planets are assumed to have no intrinsic magnetic field which could protect their atmospheres. The stellar magnetic field produces small, induced planetary magnetic fields, but the upper atmospheres of the planets are still subject to erosion. The model presented in \cite{Dong_18} captures photoionization, ion-neutral chemistry, and electron recombination chemistry that occurs in the upper atmosphere, above the neutral atmosphere dominated by $\mathrm{CO}_2$. The model then solves the MHD equations for the 4 ion fluids that exist in this case: $\mathrm{H}^+$, $\mathrm{O}^+$, $\mathrm{O}_2^+$, and $\mathrm{CO}_2^+$. The physical processes that are captured by solving these equations include stellar wind charge exchange, ion pickup, and ion sputtering, all which act to remove atmospheric material by accelerating it away from the planet. Therefore, \cite{Dong_18} is able to calculate the ion escape rates from the TRAPPIST-1 planets, which we add and use here as a total mass-loss rate in kg/s. Ion escape rates are calculated at the point where the stellar wind produces the maximum total pressure at the planets.

Given a stellar mass-loss rate from TRAPPIST-1 of $\dot{M}_{\star} = 2.6\cdot10^{8}$ kg/s, the atmospheric loss rate from TRAPPIST-1c is calculated to be $\dot{M}_\mathrm{atm} = 57$ kg/s. \cite{Dong_18} finds that the atmospheric loss rate numerically calculated in their simulations scales with planet radius, semi-major axis, and stellar mass-loss rate as follows:
\begin{equation}
\label{eq:M_dot_atm}
    \dot{M}_\mathrm{atm} \propto \left ({\frac{R_p}{a}}\right )^2\dot{M}_{\star}
\end{equation}
This scaling relation allows us to test different stellar mass-loss rates from the literature. For example, \cite{Garraffo_17} cites the stellar mass-loss rate of TRAPPIST-1 as $\dot{M}_{\star} = 1.89\cdot10^{9}$ kg/s, almost an order of magnitude higher than the value used by \cite{Dong_18}. This produces an atmospheric loss rate of $\dot{M}_\mathrm{atm} = 421$ kg/s.

These instantaneous mass-loss rates do not take into account the change in the stellar wind over time. We estimate this by connecting several M dwarf empirical relationships. As previously described, the age, rotation, high-energy radiation and mass-loss rates of stars are thought to be intimately connected. In fact, theoretical and observational work has been done to model the stellar mass-loss rate of M dwarfs as a function of age based on their X-ray luminosities and rotation periods. We employ scaling relations from \cite{Engle_18} and \cite{Magaudda_20} to calculate rotation period and X-ray luminosity, respectively:
\begin{equation}
\label{eq:P_rot}
    P_{\mathrm{rot}}[\mathrm{d}] = \frac{\mathrm{age[Gyr]} - 0.012\pm0.221}{0.061\pm0.002}
\end{equation}
\begin{equation}
\label{eq:L_X}
    L\mathrm{_X[erg/s]} = 
    \begin{cases}
        C_\mathrm{sat}P_\mathrm{rot}^{\beta_\mathrm{sat}}, & {P_\mathrm{rot}}\leq{P_\mathrm{rot,sat}}\\
        C_\mathrm{unsat}P_\mathrm{rot}^{\beta_\mathrm{unsat}}, & {P_\mathrm{rot}}>{P_\mathrm{rot,sat}}\\
    \end{cases}
\end{equation}
where $C_n = (L_{\mathrm{X},n}/P_\mathrm{rot}^{\beta_n})$ with n=(sat,unsat), $\beta_\mathrm{sat}=-0.19\pm0.11$, $\beta_\mathrm{unsat}=-3.52\pm0.02$, $P_{\mathrm{rot,sat}}=33.7\pm4.5$ [d], and log$(L_\mathrm{X,sat}(P_\mathrm{rot}=1 [\mathrm{d}]))=28.54\pm0.20$ [erg/s].
The stellar mass-loss rate can then be calculated from the X-ray surface flux as follows \citep{Wood_21}:
\begin{equation}
\label{eq:M_dot_star}
    \frac{\dot{M_\star}}{\mathrm{SA}}\left[\frac{\dot{M}_{\odot}}{\mathrm{SA_\odot}}\right] = 10^{(0.77\pm0.04\cdot\mathrm {log}_{10}(\mathrm{F_{X,surface}})-3.42)}
\end{equation}
where SA is the surface area of the star. Note that Equation \ref{eq:P_rot} does not correctly predict the known rotation period of TRAPPIST-1 at its current age. These scaling relations only provide estimates for the evolution of these quantities for a representative sample of M dwarfs, and there is significant dispersion for individual stars.

\begin{figure}
    \centering
    \includegraphics[width=3.4in]{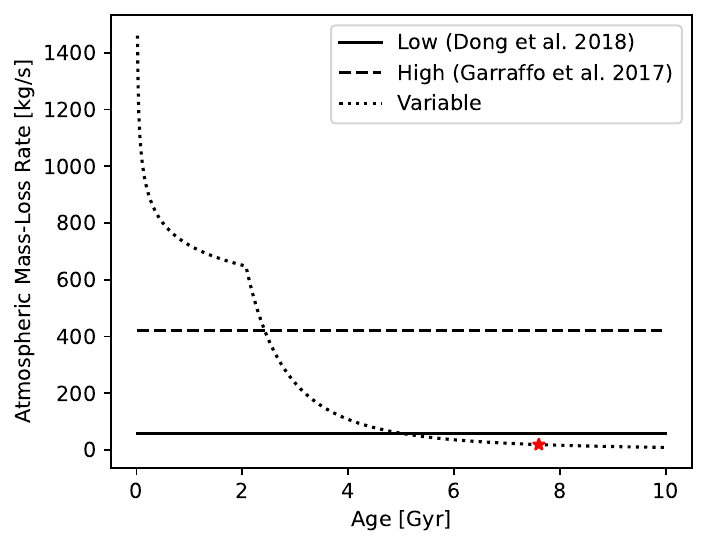}
    \caption{Three different planetary atmospheric loss rate prescriptions as a function of time used in this work. The \textit{Low} rate is taken from \cite{Dong_18}, the \textit{High} rate is derived from \cite{Garraffo_17} using Equation \ref{eq:M_dot_atm}, and the \textit{Variable} rate is calculated from M dwarf scaling relations referenced in Section \ref{sec:Atmospheric-Escape}.}
    \label{fig:Atmospheric-dMdt-Vs-Age}
\end{figure}

In this work, we use three different atmospheric loss rate prescriptions: one based on the stellar mass-loss rate presented in \cite{Dong_18} (\textit{Low}), another based on the stellar mass-loss rate presented in \cite{Garraffo_17} (\textit{High}), and a variable stellar mass rate based on the M dwarf scaling relations (\textit{Variable}). These three different cases are shown in Figure \ref{fig:Atmospheric-dMdt-Vs-Age}.

\subsection{Simulations}
\label{sec:Simulations}

We ran 2,430,000 total simulations varying atmospheric loss rate prescription (\textit{Low}, \textit{High}, and \textit{Variable}) and four key geological parameters, 3 of which are initial conditions for TRAPPIST-1c after the hydrodynamic escape phase of its evolution: 
\begin{itemize}
    \item $\mathrm{C_{tot}}$: \textbf{Initial carbon budget of the planet}. This is the amount of $\mathrm{CO}_2$ in moles that the planet initially possesses after magma ocean solidification and atmospheric desiccation, \textit{all located in the mantle} unless otherwise specified. Earth's estimated budget is $10^{22}$ mol per \cite{Foley_18}. Venus' $\mathrm{C_{tot}}$ must be at least $10^{22}$ mol, given that is approximately how much exists in the atmosphere at present.
    \item HPE: \textbf{Initial radiogenic heat-producing element} ($\mathrm{U}^{238}$, $\mathrm{U}^{235}$, K, and Th) \textbf{budget in the mantle.} These elements act to heat the mantle and may prolong outgassing.
    \item $\mu_\mathrm{ref}$: \textbf{Mantle reference viscosity.} This measures the convecting mantle material's resistance to flow, specifically at a mantle temperature equal to that of present-day Earth.
    \item $T_\mathrm{init}$: \textbf{Initial mantle temperature.} This is the starting temperature of the mantle before significant cooling occurs and depends on the planet's initial thermal budget which may come from heat from formation, impacts, etc.
\end{itemize}
Our parameter space grid comprised 30 different values of each of the four geological parameters evenly chosen from the following distributions:

\begin{equation}
    \log_{10}\left(\frac{\mathrm{C}_\mathrm{tot}}{\mathrm{C}_{\mathrm{tot},\oplus}}\right) \sim \mathscr{U}\left(\log_{10}(0.025), \log_{10}(2.5)\right)
\end{equation}
\begin{equation}
    \frac{\mathrm{HPE}}{\mathrm{HPE_{\oplus}}} \sim \mathscr{U}\left(0.5, 2\right)
\end{equation}
\begin{equation}
    \log_{10}\left(\frac{\mu_\mathrm{ref}}{\mathrm{Pa}\cdot\mathrm{s}}\right) \sim \mathscr{U}\left(20, 22\right)
\end{equation}
\begin{equation}
    \frac{T_\mathrm{init}}{\mathrm{K}} \sim \mathscr{U}\left(1700, 2000\right)
\end{equation}
Here $\mathrm{C}_{\mathrm{tot},\oplus}=10^{22}$ per an estimate of Earth's mantle and surface reservoirs from \cite{Sleep_01}. Our chosen range for $\mathrm{C}_{\mathrm{tot}}$ represents a range of $10^{-4}$ to $10^{-2}$ wt\% concentration of $\mathrm{CO}_2$ in the mantle. We test a range of initial radiogenic heat-producing element budget that spans 50\% to 200\% of Earth's budget, about 80 to 320 TW, estimated from stellar abundance measurements of radionuclides \citep{Unterborn_15, Botelho_19}. Our range of mantle reference viscosity encompasses the range of typical estimates for Earth's value. Finally, we consider a wide range of initial mantle temperature which is relatively unconstrained for Earth.

\section{Results}
\label{sec:Results}

\subsection{Early Hydrodynamic $\mathrm{CO}_2$ Loss}
\label{sec:Early-Hydrodynamic-CO2-Loss}

During the hydrodynamic hydrogen loss phase, we emphasize that it is possible for $\mathrm{CO}_2$ and other components to be dragged along with the hydrodynamic wind due to the high XUV flux of the star \citep{Fleming_20}. Here we estimate the length of the hydrodynamic H loss phase and the amount of $\mathrm{CO}_2$ lost during it per \cite{Odert_18}. For a hydrogen-dominated atmosphere with a $\mathrm{CO}_2$ component, the XUV-driven hydrodynamic flux of hydrogen in $\mathrm{kg}/\mathrm{s/m}^2$ is given by:
\begin{equation}
\label{eq:Hydrogen flux}
    \mathrm{F_H} = \frac{\beta^2 \eta \mathrm{F_{XUV}}}{4 \Delta\Phi (m_\mathrm{H} + m_{\mathrm{CO}_2} f_{\mathrm{CO}_2} x_{\mathrm{CO}_2})}
\end{equation}
Here $\beta=\mathrm{R_{XUV}}/\mathrm{R_p}$ is the ratio of the effective radius out to which the bulk of the XUV radiation is absorbed to the radius of the planet, $\eta$ is the efficiency of heating, $\mathrm{F_{XUV}}$ is the XUV flux, $\Delta\Phi=\mathrm{GM_p}/\mathrm{R_p}$, $m_\mathrm{H}$ is the mass of a hydrogen atom, $m_{\mathrm{CO}_2}$ is the mass of a $\mathrm{CO}_2$ molecule, $f_{\mathrm{CO}_2}$ is the mixing ratio of $\mathrm{CO}_2$ to H, and $x_{\mathrm{CO}_2}$ is the fractionation factor of $\mathrm{CO}_2$. The fractionation factor $x_{\mathrm{CO}_2}$ is calculated by the following equation:
\begin{equation}
\label{eq:Carbon dioxide fractionation}
    x_{\mathrm{CO}_2} = 1 - \frac{g(m_\mathrm{H} - m_{\mathrm{CO}_2}) b(T)}{\mathrm{F_H} k_B T (1 + f_{\mathrm{CO}_2})}
\end{equation}
where $b$ is the binary diffusion parameter and $T$ is the upper atmosphere temperature. $b = 8.4 \times 10^{19}  T^{0.6} \ \mathrm{m}^{-1} \ \mathrm{s}^{-1}$ for H and $\mathrm{CO}_2$. The flux of $\mathrm{CO}_2$ hydrodynamically dragged is given by:
\begin{equation}
\label{eq:Carbon dioxide flux}
    \mathrm{F_{CO_2}} = \mathrm{F_H} f_{\mathrm{CO}_2} x_{\mathrm{CO}_2}
\end{equation}
In order to track the amount of H and $\mathrm{CO}_2$ over time, Equation \ref{eq:Hydrogen flux} and \ref{eq:Carbon dioxide fractionation} must first be solved for $\mathrm{F_H}$. Then, this equation, along with Equation \ref{eq:Carbon dioxide flux}, must be integrated using a differential equation solver. When performing these calculations, we assumed $\beta = 1$, $\eta = 0.15$ \citep{Salz_15, Kubyshkina_18}, and $\mathrm{F_{XUV}}$ is the sum of EUV and X-ray flux:
\begin{equation}
\label{eq:XUV flux}
    \mathrm{F_{XUV}} = \mathrm{F_{EUV}} + \mathrm{F_{X}}
\end{equation}
We calculate X-ray flux as in Equation \ref{eq:L_X} and EUV flux per \cite{Sreejith_20} and \cite{Boudreaux_22}. We varied $T$ between 300 and 8000 K \citep{Kulikov_06} and found that it had minimal effect on the outcome. Finally, we calculated the amount of hydrogen that could be derived from a certain number of Earth oceans, and varied the number of Earth oceans (EOs) between 0.01 to 100. This initial amount of water had the greatest effect on our calculations. If TRAPPIST-1c started with 0.01 EOs, its hydrodynamic loss phase could have lasted for $\sim$ 1000 years and it could have lost about 0.1 bar of $\mathrm{CO}_2$. Starting with 1 EO, the hydrodynamic phase could last for about 100 thousand years, and about 10 bar of $\mathrm{CO}_2$ could be lost. With 100 EOs, the hydrodynamic phase could last for $\sim$ 10 Myr, and about 1000 bar of $\mathrm{CO}_2$ could be lost. We see that the length of the hydrodynamic phase is extremely short compared to the lifetime of TRAPPIST-1c in all cases, but the amount of $\mathrm{CO}_2$ lost during this phase can vary widely.

\subsection{Evolution Over Parameter Space}
\label{sec:Evolution-Over-Parameter-Space}

After calculating the rapid hydrodynamic $\mathrm{CO}_2$ loss, we focus on the $\mathrm{CO}_2$ evolution over the rest of TRAPPIST-1c's lifetime. In our long-term, coupled simulations, we track several different quantities that vary as a function of time, including mantle temperature, total amount of $\mathrm{CO}_2$ in the mantle, thickness of the lithosphere, and heat production from the radiogenic elements in the crust and mantle. Importantly, the observable quantity that we track is the amount of $\mathrm{CO}_2$ present in the atmosphere, which can be represented by mass in kg or surface pressure (p$\mathrm{CO}_2$) in bar. 

\begin{figure*}
    \centering
    \includegraphics[width=6.8in]{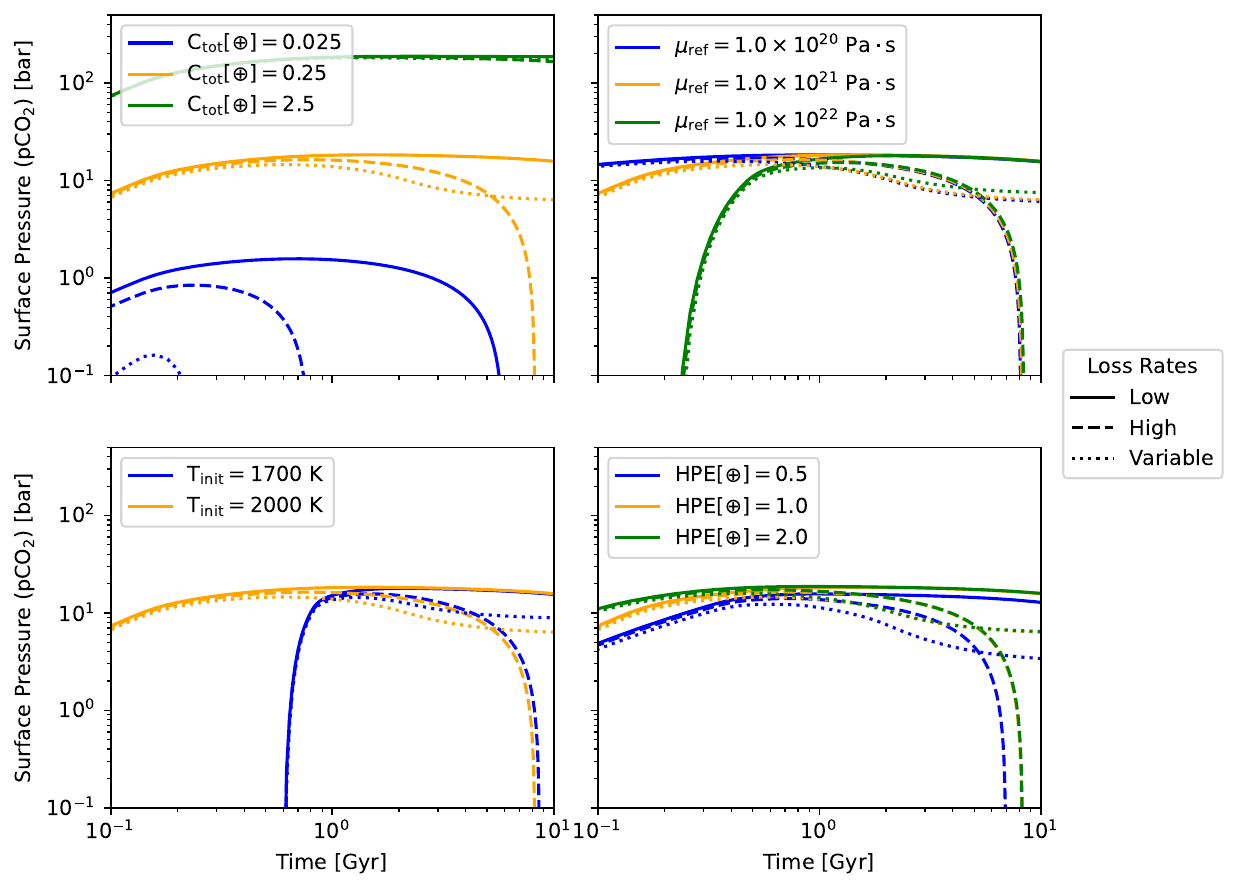}
    \caption{Evolution of atmospheric surface pressure with time in edge cases of geological parameter space with three different atmospheric loss rate prescriptions. The baseline parameters in each panel are $\mathrm{C_{tot}}[\oplus] = 0.25$, $\mu_\mathrm{ref}=1.0 \times 10^{21} \ \mathrm{Pa} \cdot \mathrm{s}$, $\mathrm{T_{init}}=2000 \ \mathrm{K}$ and $\mathrm{HPE}[\oplus]=1$.}
    \label{fig:pCO2-Vs-Time}
\end{figure*}

The modeled evolution of p$\mathrm{CO}_2$ on TRAPPIST-1c is shown for the edge cases of our varied parameters in Figure \ref{fig:pCO2-Vs-Time}. The baseline parameters are $\mathrm{C_{tot}}[\oplus] = 0.25$, $\mu_\mathrm{ref}=1.0 \times 10^{21} \ \mathrm{Pa} \cdot \mathrm{s}$, $\mathrm{T_{init}}=2000 \ \mathrm{K}$ and $\mathrm{HPE}[\oplus]=1$ in each panel. All simulations are characterized by a sharp increase in p$\mathrm{CO}_2$ in the first 1 Gyr of evolution due to early vigorous outgassing, as shown in Figure \ref{fig:Outgassing-Rate-Vs-Time}. Some simulations result in p$\mathrm{CO}_2$ that is relatively constant over the rest of TRAPPIST-1c's lifetime; others result in order of magnitude losses in p$\mathrm{CO}_2$. Specifically, the \textit{Low} atmospheric loss rate prescription results in a minimal decrease in p$\mathrm{CO}_2$ in most cases. The \textit{High} prescription leads to significant decreases in p$\mathrm{CO}_2$ that deplete the atmosphere by TRAPPIST-1's age in most cases. The \textit{Variable} prescription produces p$\mathrm{CO}_2$ curves that decrease slightly in the 1-6 Gyr range but then level out as stellar activity decreases.

The four geological parameters each have some noticeable effect on the evolution of p$\mathrm{CO}_2$. Most clearly, the initial carbon budget $\mathrm{C_{tot}}$ has a significant effect on the value of p$\mathrm{CO}_2$ that results from early outgassing. For example, $\mathrm{C_{tot}}=2.5\cdot\mathrm{C}_{\mathrm{tot},\oplus}$ leads to p$\mathrm{CO}_2$ almost as high as 200 bar, but with $\mathrm{C_{tot}}=0.025\cdot\mathrm{C}_{\mathrm{tot},\oplus}$, p$\mathrm{CO}_2$ does not even reach 2 bar. We see that in the $\mathrm{C_{tot}}=2.5\cdot\mathrm{C}_{\mathrm{tot},\oplus}$ case, none of the atmospheric loss rate prescriptions lead to a depleted atmosphere, whereas for $\mathrm{C_{tot}}=0.025\cdot\mathrm{C}_{\mathrm{tot},\oplus}$, the atmosphere is depleted before 6 Gyr for all atmospheric loss rate prescriptions. A mixture of outcomes occurs for $\mathrm{C_{tot}}=0.25\cdot\mathrm{C}_{\mathrm{tot},\oplus}$ depending on atmospheric loss rate prescription. 

HPE has a moderate effect on p$\mathrm{CO}_2$ evolution. Earth-like and twice Earth HPE budgets produce similar outcomes, while a half Earth HPE budget leads to slightly delayed outgassing and lower p$\mathrm{CO}_2$ values throughout time. $\mu_\mathrm{ref}$ and $T_\mathrm{init}$ each have minimal effects on the final p$\mathrm{CO}_2$ value at TRAPPIST-1's current age but high $\mu_\mathrm{ref}$ values and low $T_\mathrm{init}$ values do noticeably delay the onset of outgassing.

\subsection{Observational Constraints on Evolution}
\label{sec:Observational-Constraints-on-Evolution}

\begin{figure}
    \centering
    \includegraphics[width=0.5\textwidth]{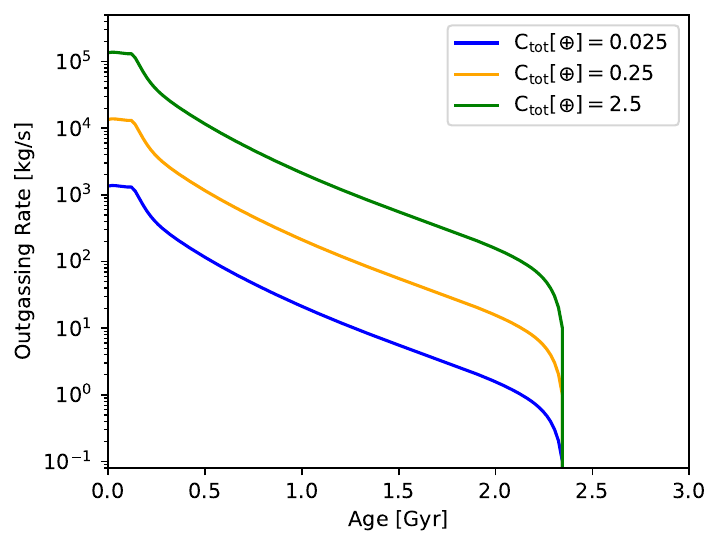}
    \caption{Outgassing rate as a function of age for three different carbon budgets $\mathrm{C_{tot}}$. Other parameters are set at $\mu_\mathrm{ref}=1.0 \times 10^{21} \ \mathrm{Pa} \cdot \mathrm{s}$, $\mathrm{T_{init}}=2000 \ \mathrm{K}$ and $\mathrm{HPE}[\oplus]=1$.}
    \label{fig:Outgassing-Rate-Vs-Time}
\end{figure}

\begin{figure}
    \centering
    \includegraphics[width=3.4in]{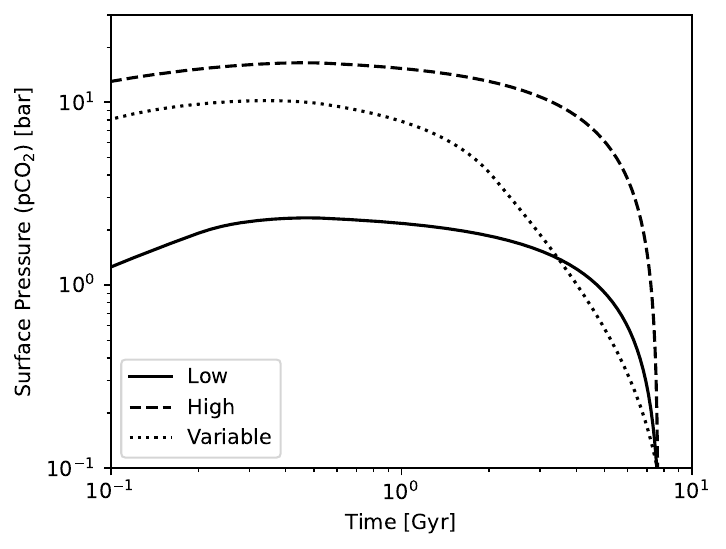}
    \caption{p$\mathrm{CO}_2$ as a function of time for the simulations, in each atmospheric loss rate prescription, that fit the observational constraint (p$\mathrm{CO}_2 < 0.1$ bar at 7.6 Gyr) \textit{and} have the highest maximum p$\mathrm{CO}_2$.}
    \label{fig:High-Max-pCO2}
\end{figure}

\cite{Zieba_23} find that TRAPPIST-1c possesses p$\mathrm{CO}_2$ $<$ 0.1 bar. This helps us constrain its atmospheric history over its $7.6\pm2.2$ Gyr lifetime, particularly after it lost its H and other constituents through hydrodynamic escape. Two broad possibilities exist: (1) an atmosphere never accumulated or (2) a non-negligible atmosphere existed at some point but was later lost. In Figure \ref{fig:Outgassing-Rate-Vs-Time}, we see that intense outgassing occurs early in the planet's lifetime and diminishes by $\sim$ 2.5 Gyr as the mantle cools. This outgassing rate depends on geological properties of the planet, specifically the initial carbon budget $\mathrm{C_{tot}}$. Therefore, there exists a degeneracy between the atmospheric mass-loss rate and the outgassing rate when attempting to differentiate between these two possibilities. If the atmospheric mass-loss rate is low, as in our \textit{Low} models, then the outgassing rate must have been low. If the atmospheric mass-loss rate is high, then the outgassing rate could have been high or low. In Figure \ref{fig:High-Max-pCO2}, we present p$\mathrm{CO}_2$ as a function of time for the simulations, in each atmospheric loss rate prescription, that fit the observational constraint \textit{and} have the highest maximum p$\mathrm{CO}_2$. In the \textit{High} models, we see that the highest p$\mathrm{CO}_2$ value that may exist in our simulations of TRAPPIST-1c is $\sim$ 16 bar. We note that this limit is relevant for the period of TRAPPIST-1c's lifetime after it became desiccated. TRAPPIST-1c could have lost a non-negligible amount of $\mathrm{CO}_2$ through drag during the hydrodynamic H loss phase as shown in Section \ref{sec:Early-Hydrodynamic-CO2-Loss}.

\begin{figure*}
    \centering
    \includegraphics[width=6.8in]{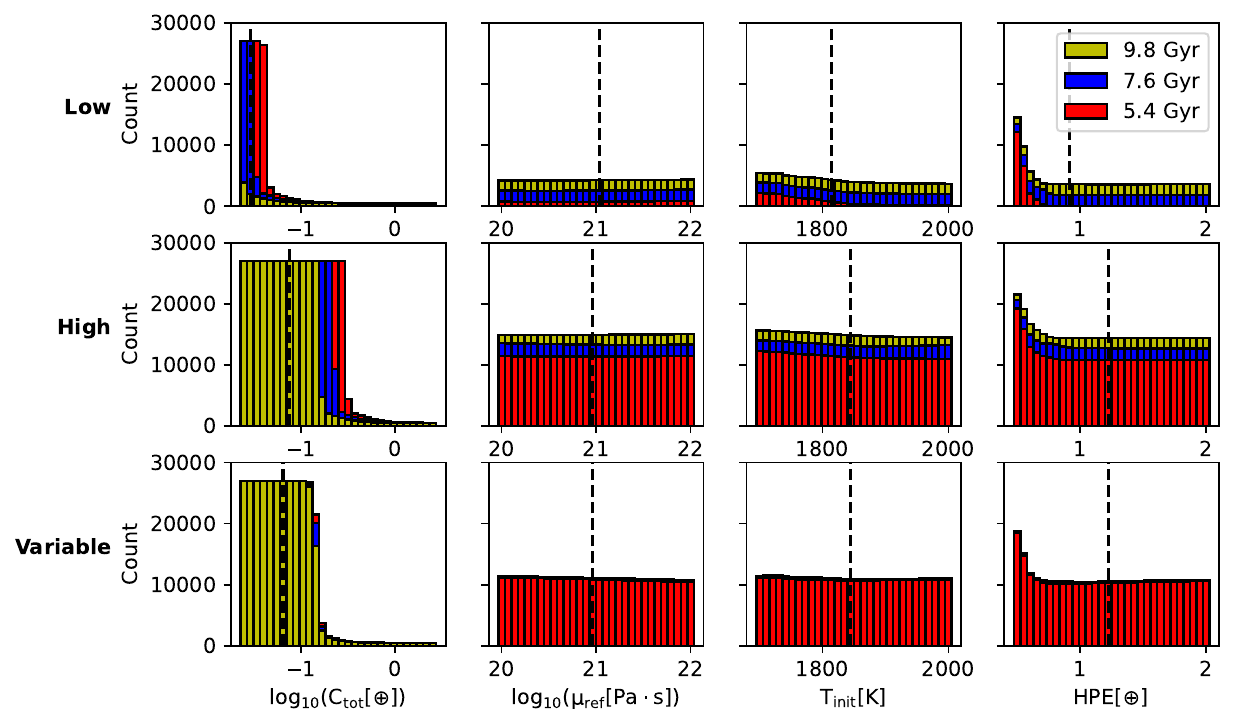}
    \caption{Histograms of parameter values of simulations (in each atmospheric loss rate prescription) that result in p$\mathrm{CO}_2$ $<$ 0.1 bar at TRAPPIST-1's measured age, 7.6 Gyr, shown in blue. The error in TRAPPIST-1's age, $\pm2.2$ Gyr, translates to some variation in these histograms which is shown in yellow and red. The median of each parameter distribution that results in p$\mathrm{CO}_2$ $<$ 0.1 bar at 7.6 Gyr is plotted as a vertical dashed line.}
    \label{fig:Successful-Simulation-Parameters-Histogram}
\end{figure*}

We now examine the values of each parameter that produced nominally ``successful'' models, or models that ended with p$\mathrm{CO}_2$ $<$ 0.1 bar at TRAPPIST-1's age, $7.6\pm2.2$ Gyr. First, note that each atmospheric loss rate prescription produced different total numbers of successful models as visualized in Figure \ref{fig:Successful-Simulation-Parameters-Histogram}. 0.095\% of \textit{Low} models were successful, 0.495\% of \textit{High} models were successful, and 0.408\% of \textit{Variable} models were successful. Assuming our parameter space encompasses all possible values of the four geological parameters and that each combination of the four parameters are equally likely in reality, this provides evidence that a stellar mass-loss-rate function similar to that modeled in \cite{Garraffo_17} or in our variable function is more likely for TRAPPIST-1 than that from \cite{Dong_18}. Within the $\mathrm{C_{tot}}$ parameter space, lower values are preferred in all atmospheric loss rate prescriptions. For \textit{Low}, \textit{High}, and \textit{Variable}, respectively, the median $\mathrm{C_{tot}}$ values are $10^{20.47}$, $10^{20.88}$ and $10^{20.81}$ moles. These are all less than Earth's estimated carbon budget $\mathrm{C_{tot,\oplus}} \sim 10^{22}$ moles (which is also Venus' lower limit). In fact, while 100$\%$ of \textit{Low} simulations with $\mathrm{C_{tot}}$ of $10^{20.47}$ are successful, this percentage drops to 18$\%$ for $\mathrm{C_{tot}}$ of $10^{20.54}$, and down to 8$\%$ for $\mathrm{C_{tot}}$ of $10^{20.60}$. Similar drop-offs in likelihood occur in \textit{High} and \textit{Variable} simulations at $\mathrm{C_{tot}}$ of $\sim10^{21.36}$ and $\mathrm{C_{tot}}$ of $\sim10^{21.23}$, respectively. The median HPE values are 0.914, 1.22, and 1.22 $\mathrm{HPE}_\oplus$ for \textit{Low}, \textit{High}, and \textit{Variable}, respectively. The distributions of successful models in $\mu_\mathrm{ref}$ and $T_\mathrm{init}$ are relatively uniform, with slightly lower values preferred for $T_\mathrm{init}$ in the \textit{Low} atmospheric loss rate prescription.

\begin{figure}
    \centering
    \includegraphics[width=3.4in]{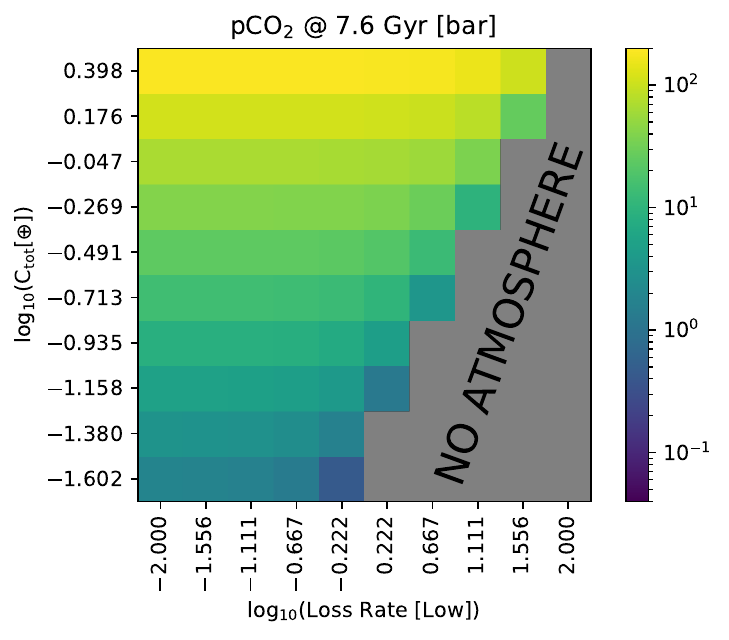}
    \caption{Atmospheric surface pressure as a function of carbon budget and loss rate, assuming loss rate is constant over time.}
    \label{fig:pCO2-Vs-Carbon-Budget-Vs-Loss-Rate-Constant}
\end{figure}

\begin{figure*}
    \centering
    \includegraphics[width=6.8in]{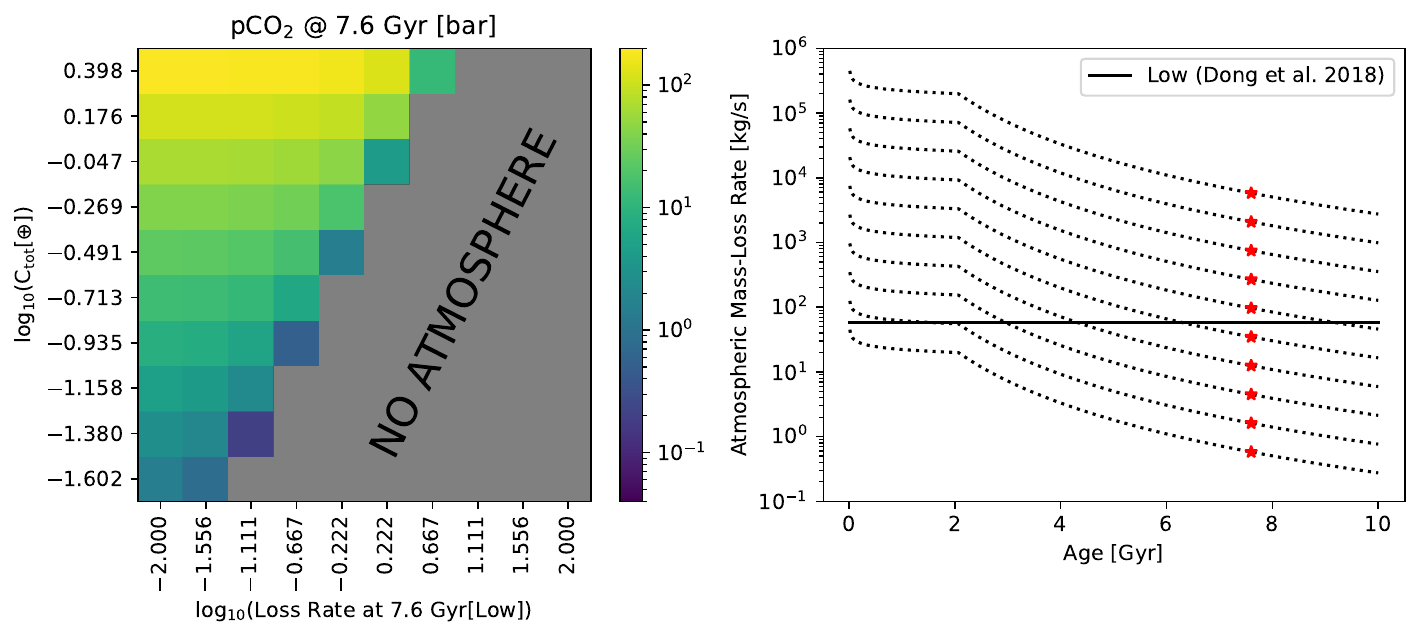}
    \caption{Atmospheric surface pressure as a function of carbon budget and reference loss rate at 7.6 Gyr, assuming loss rate is variable (left). Atmospheric mass-loss rate versus time is shown for the 10 variable cases (right). Note that the variable function is multiplied by a single factor rather than translated.}
    \label{fig:pCO2-Vs-Carbon-Budget-Vs-Loss-Rate-Variable}
\end{figure*}

Given that atmospheric loss rate, carbon budget, and p$\mathrm{CO}_2$ of TRAPPIST-1c all have uncertain values at present, we ran additional simulations varying both the atmospheric loss rate and carbon budget. We chose 10 values of $\mathrm{C_{tot}}$ uniformly log-spaced over the previously used range and 10 values of constant atmospheric loss rate uniformly log-spaced between 1\% and 10000\% of the constant \textit{Low} value. The resulting p$\mathrm{CO}_2$ values at TRAPPIST-1's age for these 100 combinations are shown in Figure \ref{fig:pCO2-Vs-Carbon-Budget-Vs-Loss-Rate-Constant}. We repeated this for variable atmospheric loss rate of the form in \textit{Variable} but with different reference loss rates at 7.6 Gyr distributed as in the previous case. This possibly more realistic set of simulations are shown in Figure \ref{fig:pCO2-Vs-Carbon-Budget-Vs-Loss-Rate-Variable}. Given that any two of atmospheric loss rate, carbon budget, and p$\mathrm{CO}_2$ at 7.6 Gyr are well known, the third quantity could be constrained using these plots.

\section{Discussion}
\label{sec:Discussion}

\subsection{Scaling Our Results for TRAPPIST-1d, e, f, g, and h}
Constraining the atmospheric evolution of TRAPPIST-1c allows us to predict the atmospheres of the other TRAPPIST-1 planets as well. Equation \ref{eq:M_dot_atm} allows us to scale the atmospheric mass-loss rate as function of semi-major axis. The outer planets have semi-major axes $a = 0.02227$, 0.02925, 0.03849, 0.04683, and 0.06189 AU, respectively. We assume that each planet has the same size as TRAPPIST-1c, with uniform volatile and radiogenic heat-producing element partitioning across the TRAPPIST-1 system, to estimate p$\mathrm{CO}_2$ for the other TRAPPIST-1 planets. 

Assuming they ``started'' out with the same atmospheric size as desiccated TRAPPIST-1c (resulting in TRAPPIST-1c having p$\mathrm{CO}_2<0.1$ bar at 7.6 Gyr), and scaling the \textit{High} atmospheric mass-loss rate, the outer planets, d, e, f, g, and h, should have less than approximately 9, 12, 15, 16, and 16 bar, respectively. The assumption that the outer planets are identically-sized to TRAPPIST-1c, is a very good approximation for TRAPPIST-1g, a fair approximation for TRAPPIST-1f, and less so for the remaining smaller outer planets. These calculations suggest that the 5 outer planets in the TRAPPIST-1 system may possess atmospheres even if TRAPPIST-1c has little atmospheric material. Further, the assumption that the outer planets had the same amount of atmospheric material as TRAPPIST-1c after it became desiccated is also likely not a good approximation, because the process of runaway greenhouse and hydrodynamic escape would not occur to the same extent for the outer planets. This implies that they would have even larger atmospheres than calculated here. This agrees qualitatively with the recent paper by \citet{Krissansen-Totton_23}, who found that TRAPPIST-1e and f retained atmospheres in 98\% of their model runs.

\subsection{Atmospheric Processes}
\label{sec:Atmospheric-Processes}

Our simulations assume a pure $\mathrm{CO}_2$ atmosphere, which is characteristic of present-day Venus and Mars, but may not be representative of all terrestrial planet atmospheres. In fact, non-CO$_2$ atmospheres remain a viable possibility consistent with the observations for TRAPPIST-1c \citep{Zieba_23, Lincowski_23}. In the solar system, Titan possesses a $\mathrm{N}_2$-dominated atmosphere with a non-negligible amount of $\mathrm{CH}_4$. We chose a pure $\mathrm{CO}_2$ atmosphere because it is common in the Solar System, and is the highest mean molecular weight atmosphere possible. If any atmospheric material remains after intense escape, it should be $\mathrm{CO}_2$. In this sense, assuming a pure $\mathrm{CO}_2$ atmosphere represents an upper limit on the size of terrestrial planet atmospheres. A large abiotic $\mathrm{O}_2$ component may be left after the early hydrodynamic escape phase, but this will escape more easily than $\mathrm{CO}_2$. Additional species could exist, depending on available material during formation and the redox state of the planet's mantle. If we were to add more components, like $\mathrm{N}_2$ and $\mathrm{CH}_4$, we would expect these components to be lost more easily, and more complex photochemistry would operate. 

 While a large amount of water likely escapes from TRAPPIST-1c early, it is also reasonable to assume that some non-negligible amount of water (along with other trace gases) is outgassed with $\mathrm{CO}_2$. This would replenish the atmosphere with water vapor and other gases over the lifetime of TRAPPIST-1c, which would then be removed simultaneously with $\mathrm{CO}_2$. Given the same total mass-loss rate, and the observed constraint on p$\mathrm{CO}_2$, the initial carbon budget after desiccation $\mathrm{C_{tot}}$ might be even lower. Thus, the initial carbon budget after desiccation that we find is an upper bound.

Another escape mechanism that could, in theory, affect the composition and size of TRAPPIST-1c's atmosphere is impact erosion, which we do not model in this manuscript. However, \cite{Raymond_22} run N-body simulations and find that high rates of impact erosion are quite unlikely in the TRAPPIST-1 system after formation because perturbations from additional objects could break the multi-resonant orbital configuration of the system.

\subsection{Geophysical Processes}
\label{sec:Geophysical-Processes}

We have assumed that the planet operates under a stagnant-lid tectonic regime due to its lack of water \citep{Tikoo_17}, and the fact that plate tectonics are rare in the Solar System. However, TRAPPIST-1c has higher gravity than planets in the Solar System, which may increase the likelihood of subduction and, therefore, plate tectonics. If plate tectonics does operate on TRAPPIST-1c, previous work has suggested that outgassing rates may be higher for these ``active lids'' as opposed to ``sluggish'' or stagnant lids \citep{Fuentes_19}. Without water, recycling would still not occur, so it is plausible that the mantle's carbon reservoir would be depleted more quickly. Depending on the atmospheric mass-loss rates, this could result in a shorter atmospheric retention time.

\begin{figure*}
    \centering
    \includegraphics[width=6.8in]{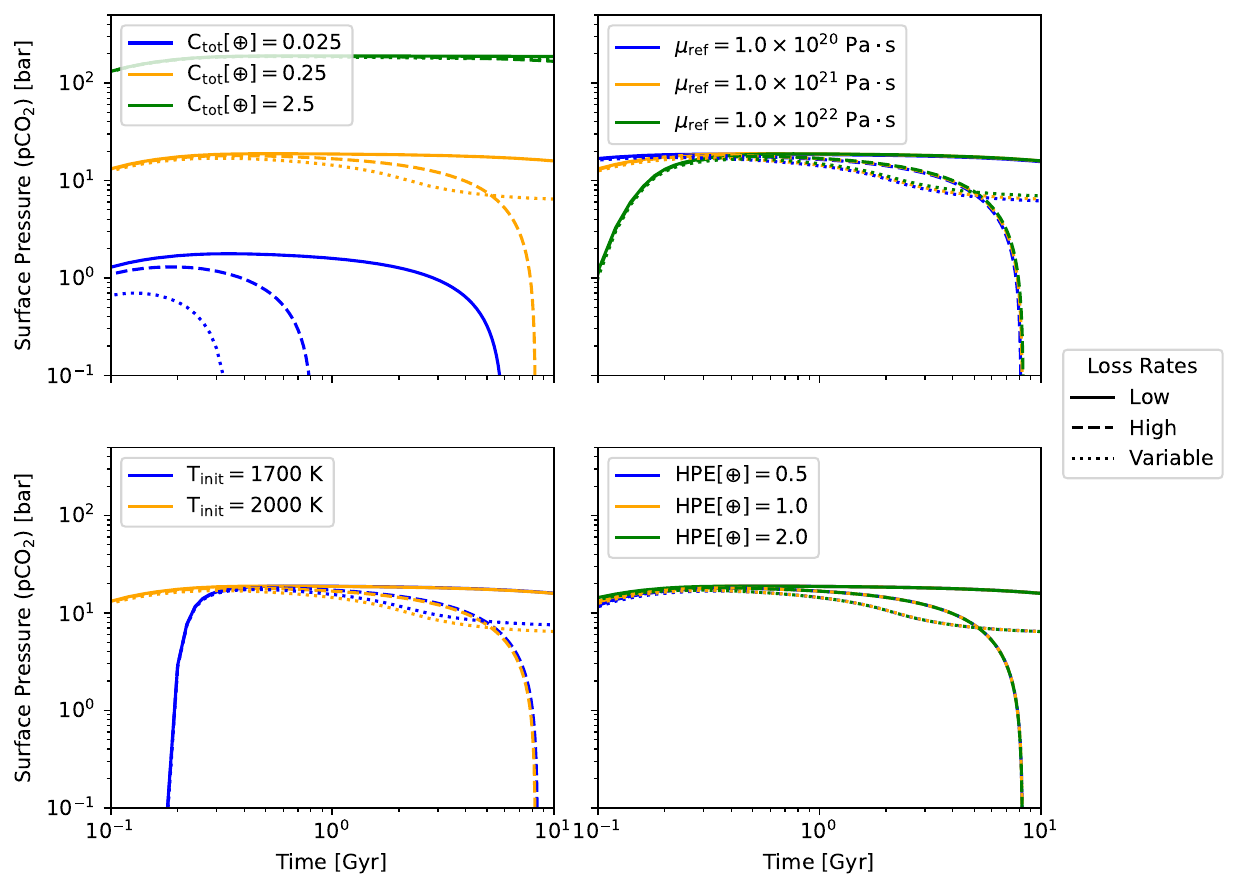}
    \caption{Evolution of atmospheric surface pressure with time in edge cases of geological parameter space with three different atmospheric loss rate prescriptions\textbf{ with constant tidal heating rate added}. The baseline parameters in each panel are $\mathrm{C_{tot}}[\oplus] = 0.25$, $\mu_\mathrm{ref}=1.0 \times 10^{21} \ \mathrm{Pa} \cdot \mathrm{s}$, $\mathrm{T_{init}}=2000 \ \mathrm{K}$ and $\mathrm{HPE}[\oplus]=1$.}
    \label{fig:pCO2-Vs-Time-w-Tidal-Heating}
\end{figure*}

Another geophysical consideration that is relevant to TRAPPIST-1c is the effect of tidal heating. Thus far, we have assumed that the only heating source in the mantle is from radiogenic elements. However, \cite{Dobos_19} found that the internal heat flux from tidal heating for TRAPPIST-1c is 0.62 $\mathrm{W/m}^2$, comparable to that observed on Io (1-2 $\mathrm{W/m}^2$). We added this a constant heating rate to our thermal evolution, which has the effect of increasing the equilibrium mantle temperature and pushing vigorous outgassing to even earlier times. We ran the edge cases of our four geological parameters again as in Figure \ref{fig:pCO2-Vs-Time}, to see how tidal heating affected the $\mathrm{pCO}_2$ evolution. The results are shown in Figure \ref{fig:pCO2-Vs-Time-w-Tidal-Heating}. Comparing to the results without tidal heating in Figure \ref{fig:pCO2-Vs-Time}, the final $\mathrm{pCO}_2$ value at TRAPPIST-1's age is relatively unchanged in each case. The largest noticeable difference is that $\mathrm{pCO}_2$ increases at earlier times in all cases, due to an earlier onset of vigorous outgassing. The specific HPE value becomes unimportant in determining the $\mathrm{pCO}_2$ evolution, because tidal heating becomes the dominant heating term. Overall, tidal heating does not change our conclusion that low initial carbon budgets are significantly preferred for TRAPPIST-1c to match observations.

\begin{figure}
    \centering
    \includegraphics[width=3.3in]{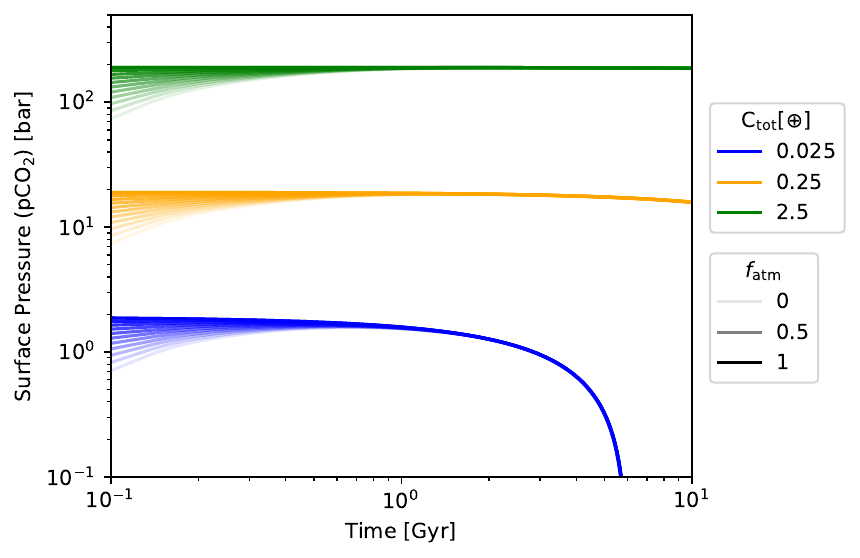}
    \caption{Evolution of atmospheric surface pressure with time, varying $\mathrm{C_{tot}}$ and $f_\mathrm{atm}$, the fraction of $\mathrm{C_{tot}}$ placed in the atmosphere. The baseline parameters are $\mu_\mathrm{ref}=1.0 \times 10^{21} \ \mathrm{Pa} \cdot \mathrm{s}$, $\mathrm{T_{init}}=2000 \ \mathrm{K}$ and $\mathrm{HPE}[\oplus]=1$.}
    \label{fig:pCO2-Vs-Time-w-Varied-C-Location}
\end{figure}

One more geophysical process that occurs during the magma ocean phase of a planet and subsequent solidification of the magma ocean is volatile partitioning. We assume in this manuscript that the initial carbon budget of the planet (after magma ocean solidification and planetary desiccation) is all located in the mantle, and it subsequently outgasses from there. However, studies of magma ocean evolution have shown that the majority of $\mathrm{CO}_2$ may partition into the atmosphere (e.g. \citealt{Hier-Majumder_17, Bower_22}). We performed additional simulations varying the fraction of $\mathrm{CO}_2$ $f_\mathrm{atm}$ starting out in the atmosphere for different initial carbon budgets $\mathrm{C_{tot}}$. In this case, $\mathrm{C_{tot}}$ still refers to the total amount that the planet possesses, but it is divided between the atmosphere and the mantle. Figure \ref{fig:pCO2-Vs-Time-w-Varied-C-Location} shows the $\mathrm{pCO}_2$ evolution from these simulations. It is clear that different $f_\mathrm{atm}$ values slightly change the evolution of $\mathrm{pCO}_2$ before 1 Gyr, but do not affect the outcome at 7.6 Gyr.

\subsection{Star-Planet Interaction}
\label{sec:Star-Planet-Interaction}

A number of assumptions introduced by the MHD models \citep{Dong_18} whose results we use in our simulations may impact our results. The ion escape rates are calculated in the case where the TRAPPIST-1 planets are unmagnetized and subject to maximum total wind pressure, as to obtain upper limits on escape rates. In particles per second, \cite{Dong_18} calculates the total ion escape rates to be on the order of $10^{27}$ for the TRAPPIST-1 planets. \cite{Dong_19} did similar simulations for the TRAPPIST-1 planets, varying the planetary obliquity, but only included $\mathrm{H}^+$ and $\mathrm{O}^+$ ions, so the ion escape rates are higher ($10^{28}$). Other recent work has calculated ion escape rates for $\mathrm{CO}_2$-containing atmospheres in the case of early Earth, and found a value of $\sim 10^{27}$/s at 4.0 Gyr ago \citep{Grasser_23}. This is similar to that found in \cite{Dong_18} for a much higher stellar wind pressure. This discrepancy can be explained by the fact that \cite{Grasser_23} modeled a $\mathrm{CO}_2$/$\mathrm{N}_2$ atmosphere which is more susceptible to escape and included XUV-induced expansion of the upper atmosphere which \cite{Dong_18} does not.

\cite{Dong_18} uses a solar magnetogram as input to their models and scales the magnetic field strength based on observations of similar late M dwarfs \citep{Morin_10} (\cite{Dong_18} don't report the exact value they use, but the average of the targets measured in \cite{Morin_10} is $\sim 500$ G). However, \cite{Garraffo_17} uses the magnetogram of M6.5 dwarf GJ3622 and 600 G for the magnetic field strength as found by Zeeman Broadening of TRAPPIST-1 \citep{Reiners_10}. These differences are likely the source of the discrepancy in stellar mass-loss rates that \cite{Dong_18} and \cite{Garraffo_17} report. Future observations should aim to better constrain these inputs. Further, the MHD models require an input of XUV luminosity to calculate photoionization and resultant stellar heating, which \cite{Dong_18} draws from \cite{Bourrier_17} 
to be $F_\mathrm{XUV}=801\substack{+436 \\ -217}$ ergs/s/$\mathrm{cm}^2$. The XUV flux of TRAPPIST-1 should also vary as a function of stellar age which is not included in the MHD models currently.

\section{Conclusions}
\label{sec:Conclusions}

In this manuscript, we have presented simulations of outgassing and escape that evolve the atmosphere of TRAPPIST-1c, assuming it is $\mathrm{CO}_2$-dominated. Given our assumptions about atmospheric escape and geological phenomena, and based on the constraint of p$\mathrm{CO}_2$ $<$ 0.1 bar from observations, we find that:
\begin{enumerate}[parsep=0pt,itemsep=0pt]
    \item Long-term stellar wind stripping is not efficient enough to remove a large $\mathrm{CO}_2$ atmosphere from TRAPPIST-1c. Specifically, 
    \begin{itemize}
        \item The median initial carbon budget after desiccation ($\mathrm{C_{tot}}$) is significantly less than modern Earth's carbon budget and Venus' lower limit, $10^{22}$ mol, in all atmospheric loss rate prescriptions.
        \item The surface pressure of $\mathrm{CO}_2$ in TRAPPIST-1c's past, following early hydrodynamic loss, could have only been as high as $\sim 16$ bar.
    \end{itemize}
    \item TRAPPIST-1c must have either formed volatile-poor as compared to Earth and Venus, or lost a substantial amount of $\mathrm{CO}_2$ during the early hydrodynamic escape phase which requires a high initial water inventory -- at least $\sim$ 10 Earth oceans.
    \item The outer TRAPPIST-1 planets may retain substantial atmospheres which are $\sim 2$ orders of magnitude larger than that of TRAPPIST-1c.
\end{enumerate}

Our results show that TRAPPIST-1c's low constraint on p$\mathrm{CO}_2$ today traces back to a lower carbon budget than Earth and Venus after desiccation. This divergence in evolution could be due to a difference in total volatile inventory (Scenario 1) which may be due to variations in protoplanetary disk composition and size, or it might relate to orbital dynamics, such as the scattering of volatile-rich planetesimals due to giant planets during formation \citep{Raymond_04}. An equally valid scenario (2) exists where TRAPPIST-1c's volatile inventory was similar to that of Earth and Venus, and the high-energy radiation environment of TRAPPIST-1 was harsh enough to deplete it in its first $10^5$--$10^7$ years.

More \textit{JWST} observations in various wavelength bands will give us further insight into the TRAPPIST-1 planets' atmospheric sizes and compositions. While we expect $\mathrm{CO}_2$ to be the most resistant species to intense escape, other species could represent signatures of atmospheric evolution. It is also plausible that TRAPPIST-1c and its neighbors do, in fact, lack atmospheres; the observations presented here from \citet{Zieba_23} only provide an upper limit. Regardless, evolutionary models, which couple stellar, atmospheric, and geological processes, like that which we have presented here, will be key to understanding how these planets became what they are today.

\section{Code Availability}
\label{sec:code-availability}
We make our code publicly available. The software is available on GitHub (\url{https://github.com/katieteixeira/atmospheric_evolution}) under a MIT License and version 1.0.0 is archived in Zenodo \citep{teixeira_&_foley_code}. It can be used to run simulations, save and load data, and make figures.

\begin{acknowledgments}

We thank Brendan Bowler, Adam Kraus, Amber Medina, David Wilson, and Cynthia Froning for fruitful discussions that aided the development of this project and manuscript. This work was supported through a Student Research Award in Planetary Habitability from the UT Center for Planetary Systems Habitability. Support for programs HST-AR-17025.001-A, HST-AR-15805.001-A, and JWST-GO-02304.002-A were provided by NASA through a grant from the Space Telescope Science Institute, which is operated by the Associations of Universities for Research in Astronomy, Incorporated, under NASA contract NAS5-26555. C.V.M. acknowledges support from the Alfred P. Sloan Foundation under grant number FG-2021-16592. 

\end{acknowledgments}

\bibliography{main.bib}

\end{document}